\documentclass[multphys,vecphys]{svmult}

\usepackage{makeidx}         
\usepackage{graphicx}        
\usepackage{multicol}        
\usepackage[bottom]{footmisc}

\makeindex             

\begin{document}

\newfont{\reiprichfont}{cmssqi8 scaled 1200}
\newcommand{\reiprichgcs}{{\reiprichfont HIFLUGCS}}

\title*{Complex Physics in Cluster Cores: Showstopper for the Use of Clusters
for Cosmology?}
\titlerunning{Complex Cluster Cool Cores and Cosmology}
\author{Thomas H. Reiprich
\and
Daniel S. Hudson
}
\institute{Argelander-Institut f\"ur Astronomie, Universit\"at Bonn, Auf dem
H\"ugel 71, 53121 Bonn, Germany
\texttt{thomas@reiprich.net, dhudson@astro.uni-bonn.de}
}
%
%
\maketitle

\section{Introduction}
\label{reiprich:intro}

Galaxy clusters can be rather messy objects, e.g.~\cite{rsk04}. Why should one
use them to help solving pressing cosmological problems, especially about the
nature of dark matter and dark energy? Are there not cleaner probes for
this purpose?

First of all, the possible implications of dark energy, e.g., a
modification of the fundamental gravity law or an introduction of a fifth
force, are too far-reaching that we could afford to rely on just one single
method: several independent observational methods are necessary if our picture
of the universe is to be changed dramatically.
Secondly, measurements of the primary anisotropies in the cosmic microwave
background are not sensitive to any evolution of the equation of state of dark
energy. Thirdly, it would appear that we may be able to simulate relevant
physical processes in galaxy clusters actually \emph{more} realistically than,
e.g., in galaxies or supernovae. That is, clusters may indeed be
\emph{relatively} simple and clean probes.  Fourthly, with purely geometric
tests, e.g., using supernovae as standard candles, we cannot differentiate
between, e.g., quintessence and a possible breakdown of general relativity.
This can, however, be achieved with tests based on structure growth, e.g., the
evolution of the galaxy cluster mass function.

Moreover, clusters are unique cosmological probes in the sense that there are
many, more or less independent methods to constrain cosmological parameters with
clusters and basically all wavelengths can be used to study clusters.
Tests include, e.g., cluster baryons (fraction and its \emph{apparent} evolution),
power spectrum (normalization, shape, and baryonic wiggles), mergers (frequency
and its evolution), and mass function (normalization, shape, and evolution).
Wavelengths to find and study clusters include, e.g., optical/infrared
(galaxies, lensing), radio (Sunyaev--Zeldovich-effect, halos and relics, wide
and narrow angle tailed galaxies), $\gamma$-rays (especially with future
instruments like GLAST), and X-rays.

Finally, after a phase of skepticism, renewed trust in clusters seems to
spread. Skepticism was in part caused by low values of $\sigma_8\sim 0.7$
(for $\Omega_{\rm m}=0.3$) indicated early from cluster studies
\cite{brt01,rb01,sel01,vnl01}, which seemed to be at variance with $\sigma_8$
values obtained from other probes, including the 1st year WMAP data
\cite{svp03}. This has changed since the release of the 3rd year WMAP data,
which now confirms the low $\sigma_8$ values \cite{sbd06}. Furthermore, the
new best fit results from WMAP indirectly suggest that the intrinsic
scatter and bias of cluster scaling relations like the X-ray
luminosity--gravitational mass ($L_{\rm X}$--$M_{\rm tot}$) relation may be
smaller than previously thought \cite{r06}.

For future determinations of the evolution of the cluster mass function
with the new generation of X-ray surveys (e.g., eROSITA is expected to detect
about 100\,000 clusters), primarily only X-ray luminosites will be available (gas
temperatures only for a small subset of clusters). Therefore, we concentrate in
this contribution on effects of cluster physics on the $L_{\rm X}$--$M_{\rm
tot}$ relation. And since this is a cooling flow conference, we concentrate on
the influence of cool cores on this relation.

\section{Cool cores and the luminosity--mass relation}
\label{reiprich:gcs}
As mentioned above, indirectly the WMAP 3rd year data require no large bias or
intrinsic scatter in the $L_{\rm X}$--$M_{\rm tot}$ relation. However, we would
rather like to determine the intrinsic scatter directly from the data. This can
be difficult because the measured scatter is a combination of statistical,
systematic, and intrinsic scatter. So, a detailed understanding of all
relevant systematic effects is required for a reliable determination of the
intrinsic scatter. We are confident that the high quality cluster samples and
state of the art data now available from Chandra, XMM-Newton, and Suzaku will be
sufficient for a good estimate. We are currently working on this using the
\reiprichgcs\ clusters. The preliminary results we show in this
contribution are very closely related to other work that has been done recently
with older data \cite{omb06,crb07}.

\begin{figure}
\centering
\includegraphics[height=6.8cm]{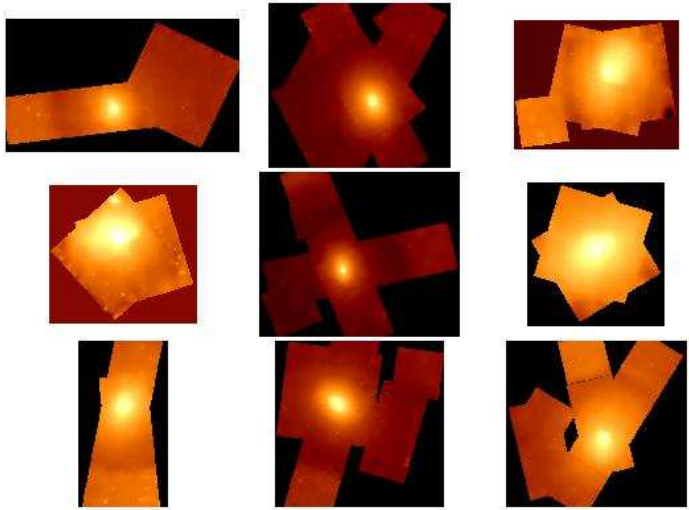}
%
%
\caption{Chandra observations for nine exemplary clusters in \reiprichgcs. All
available observations and all usable CCDs are analyzed in order to maximize
signal-to-noise ratio and field-of-view.} 
\label{reiprich:mosaics}       
\end{figure}
\reiprichgcs\ contains the 64 X-ray brightest clusters in the sky excluding
$\pm$20 deg around the Galactic plane and some small regions around the
Magellanic clouds and the Virgo cluster. It is a complete X-ray flux-limited
sample selected from deep surveys based on the ROSAT All-Sky Survey \cite{rb01} 
(RB02).
It is currently the best available sample in terms of homogeneous selection,
size, completeness, representativeness, and full Chandra and (almost) XMM-Newton 
coverage. We are currently analyzing $>$120 Chandra and $>$100 XMM-Newton
observations with a total exposure time approaching 7 Ms (see
Fig.~\ref{reiprich:mosaics}, and Hudson \& Reiprich, these proceedings, and
Nenestyan \& Reiprich, these proceedings).

We are currently studying several methods to classify clusters as cool core (CC)
and non-cool core (NCC) clusters with Chandra, including the slopes of the inner
temperature and density profiles, central cooling times (the time the gas
needs to cool below X-ray emitting temperatures), and central entropies. There
is a large but not complete overlap between the results of these methods. Here
we use a special ``central'' entropy to select CC (low entropy; i.e., high
density and low temperature) and NCC (high entropy) clusters (see Hudson \&
Reiprich, these proceedings).

\begin{figure}
\centering
\includegraphics[height=4.1cm]{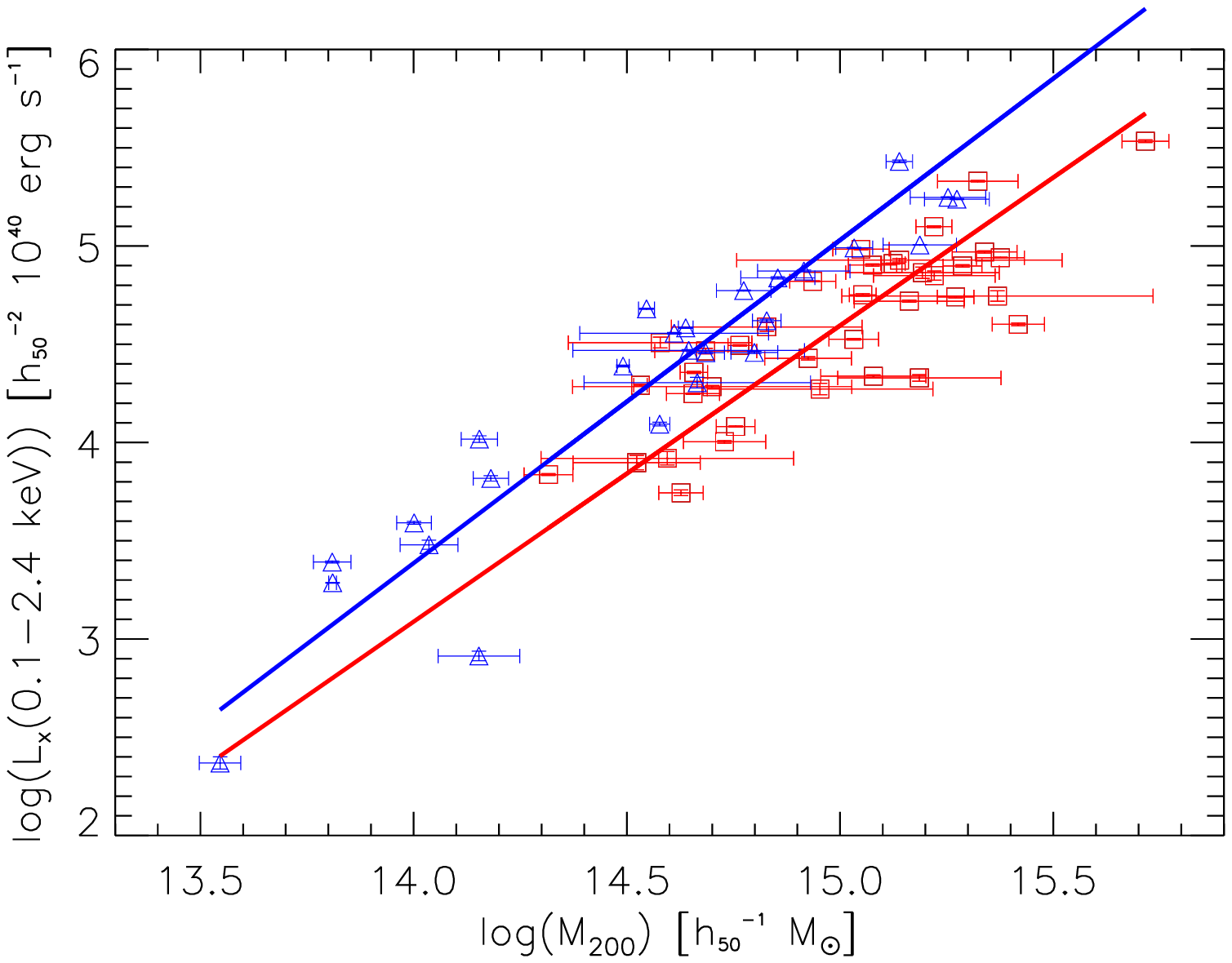}
\includegraphics[height=4.1cm]{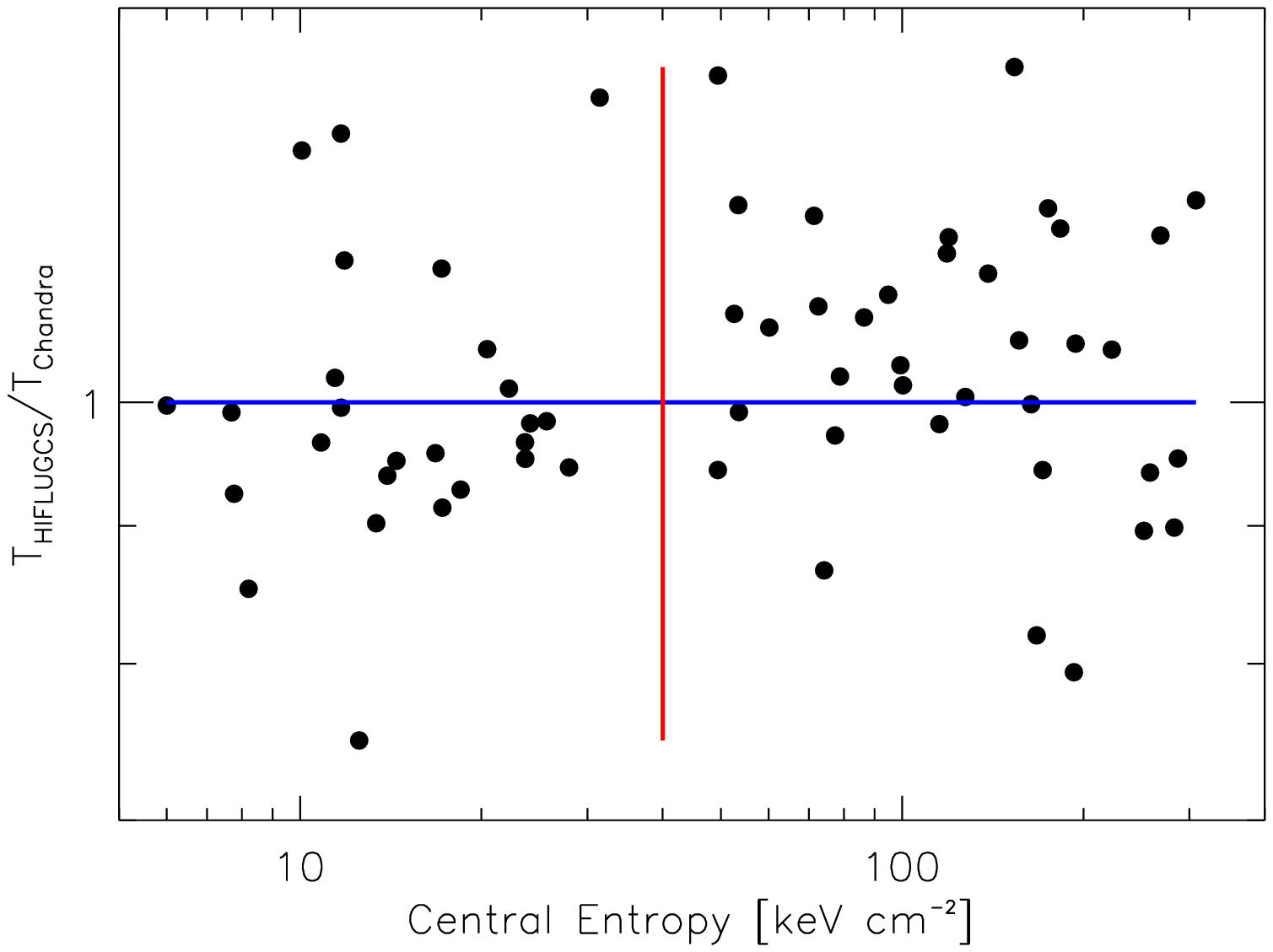}
%
%
\caption{\emph{Left:} $L_{\rm X}$ vs.\ $M_{\rm tot}$ for the \reiprichgcs\ 
clusters (RB02). Blue triangles represent CC clusters and red squares NCC
clusters as classified through the central entropy determined with Chandra. The
best fit (bisector) relations show that CC clusters have a higher normalization.
\emph{Right:} Ratio of original \reiprichgcs\ and new (preliminary) Chandra
temperatures vs.\ central entropy (CC clusters are to the left, NCC clusters to
the right). Overall, there is quite good agreement between the temperature
estimates; however, most of the CC clusters have a ratio below 1 while most of
the NCC clusters have a ratio larger than 1.}  
\label{reiprich:L-M}       
\end{figure}

Now let us check if the two populations, CC and NCC clusters, behave differently
in the $L_{\rm X}$--$M_{\rm tot}$ diagram. For the nearby clusters in
\reiprichgcs, $L_{\rm X}$ is best determined with ROSAT data due to its large
field-of-view and low background. Gravitational masses have not, yet, been
determined with Chandra or XMM-Newton for all \reiprichgcs\ clusters so we
simply use the old 
masses determined from the ROSAT gas density profiles and overall (primarily
ASCA) gas temperatures (RB02). Figure~\ref{reiprich:L-M} (left) shows that the
$L_{\rm X}$--$M_{\rm tot}$ relation for CC clusters has a factor of 2.5
higher normalization (at $5\times10^{14}M_{\odot}$) than the relation for NCC
clusters -- the CC clusters segregate out to the high $L_{\rm X}$ (or low
$M_{\rm tot}$) side (see also Chen et al., these proceedings). Also, the CC
clusters seem to exhibit smaller scatter around their best fit relation than the
NCC clusters. This may be at variance with the results of O'Hara et al.\
\cite{omb06} who found a \emph{larger} scatter for CC clusters in the $L_{\rm
X}$--$T_{\rm gas}$ relation. Furthermore, it appears that all low mass clusters
and groups in the sample have a cool core.

The factor 2.5 offset between the two best fit relations may indicate
significant intrinsic scatter; i.e., since CC clusters have higher
central densities and since X-ray emissivity is proportional to density squared,
CC clusters may have significantly higher $L_{\rm X}$ for given $M_{\rm tot}$
compared to NCC clusters, if the central regions of CC clusters account for a
very significant fraction of the total cluster luminosities.

On the other hand, systematic effects can play a role as well. If, e.g., cool
cores bias overall 
cluster temperature estimates low compared to their virial temperatures then the
estimated masses will be biased low, too. A possible mass bias is enhanced
compared to a temperature bias because $M_{\rm tot}\propto T_{\rm gas}^{1.5}$
and the offset to be accounted for in $M_{\rm tot}$ direction is smaller then
the offset in $L_{\rm X}$ direction because $M_{\rm tot}\propto L_{\rm
X}^{1.4}$. So, even relatively small $T_{\rm gas}$ biases can have a significant
effect on the $L_{\rm X}$ offset in the $L_{\rm X}$--$M_{\rm tot}$ relation.
Also other systematic differences between CC and NCC clusters with  
the potential of biasing simple mass estimates might be important, e.g., a
difference between the steepening of the surface brightness profiles in the very
outer CC/NCC cluster parts (e.g., Burns et al., these proceedings). 

Many of the temperature estimates we used for the original mass determination in
RB02 were, one way or another, ``corrected'' for cooling flows. So, we actually
do not expect a very large bias. With the new preliminary Chandra temperature
profiles for all \reiprichgcs\ clusters available (Hudson \& Reiprich, these
proceedings) it is straightforward to exclude thoroughly any cool core emission
for overall temperature estimates. Figure~\ref{reiprich:L-M} (right) shows the
ratio of the original temperature estimates and the preliminary Chandra overall
$T_{\rm gas}$ determinations as a function of central entropy. Clusters to the
left in this diagram are CC clusters, those to the right NCC clusters. While in
general there is very good agreement between the temperatures, one notes that
most of the CC clusters have a ratio below 1 while most of the NCC clusters have
a ratio larger than 1, indicating that indeed a small temperature bias is
present. However, currently the magnitude of this effect alone does not seem
large enough to account for all of the observed $L_{\rm X}$ offset. Soon the 
Chandra analysis will be completed (including the mass determination). We will
then be able to derive very tight and robust limits on the intrinsic scatter in
the $L_{\rm X}$--$M_{\rm tot}$ relation.

\begin{figure}
\centering
\includegraphics[height=4.1cm]{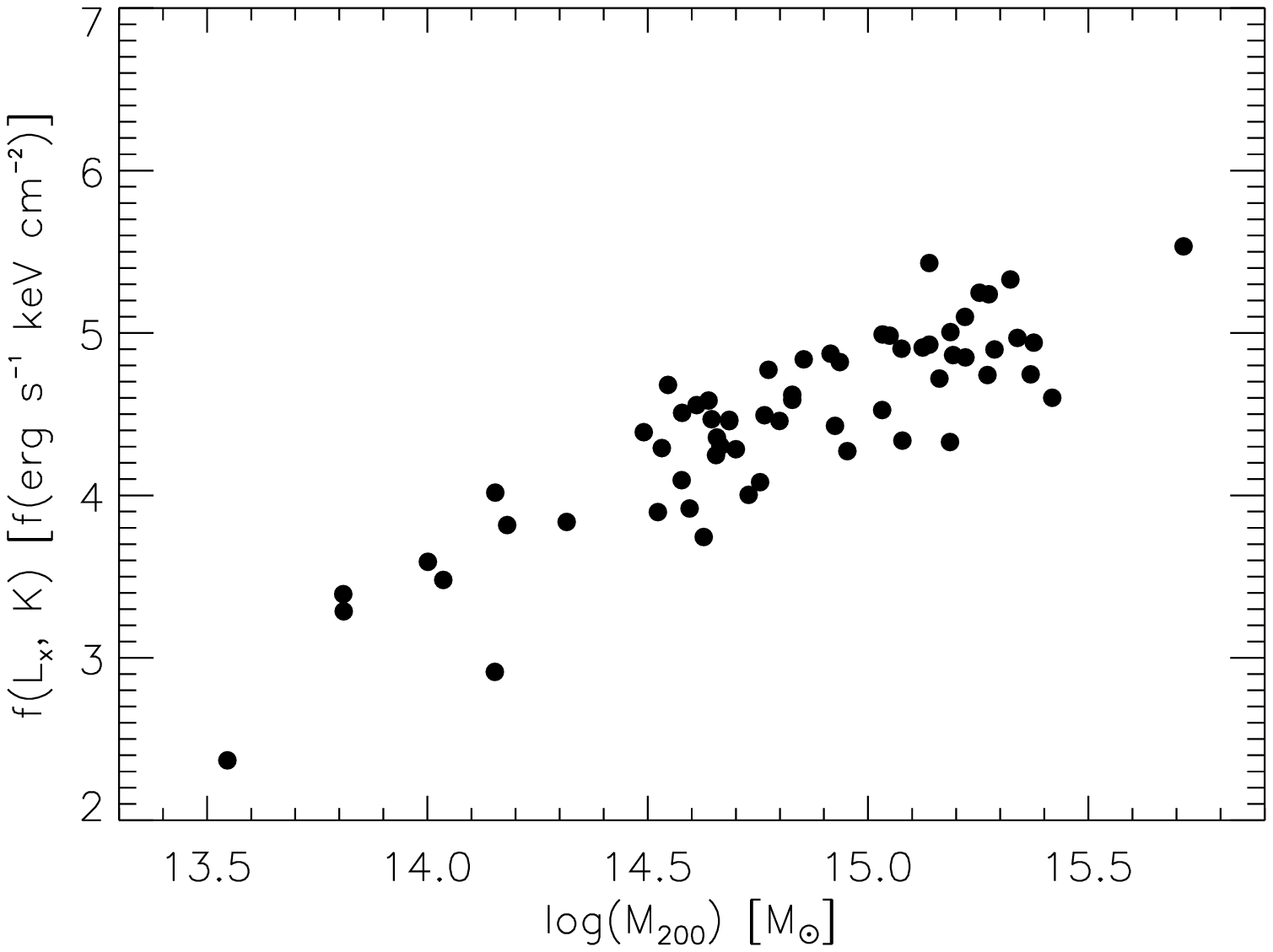}
\includegraphics[height=4.1cm]{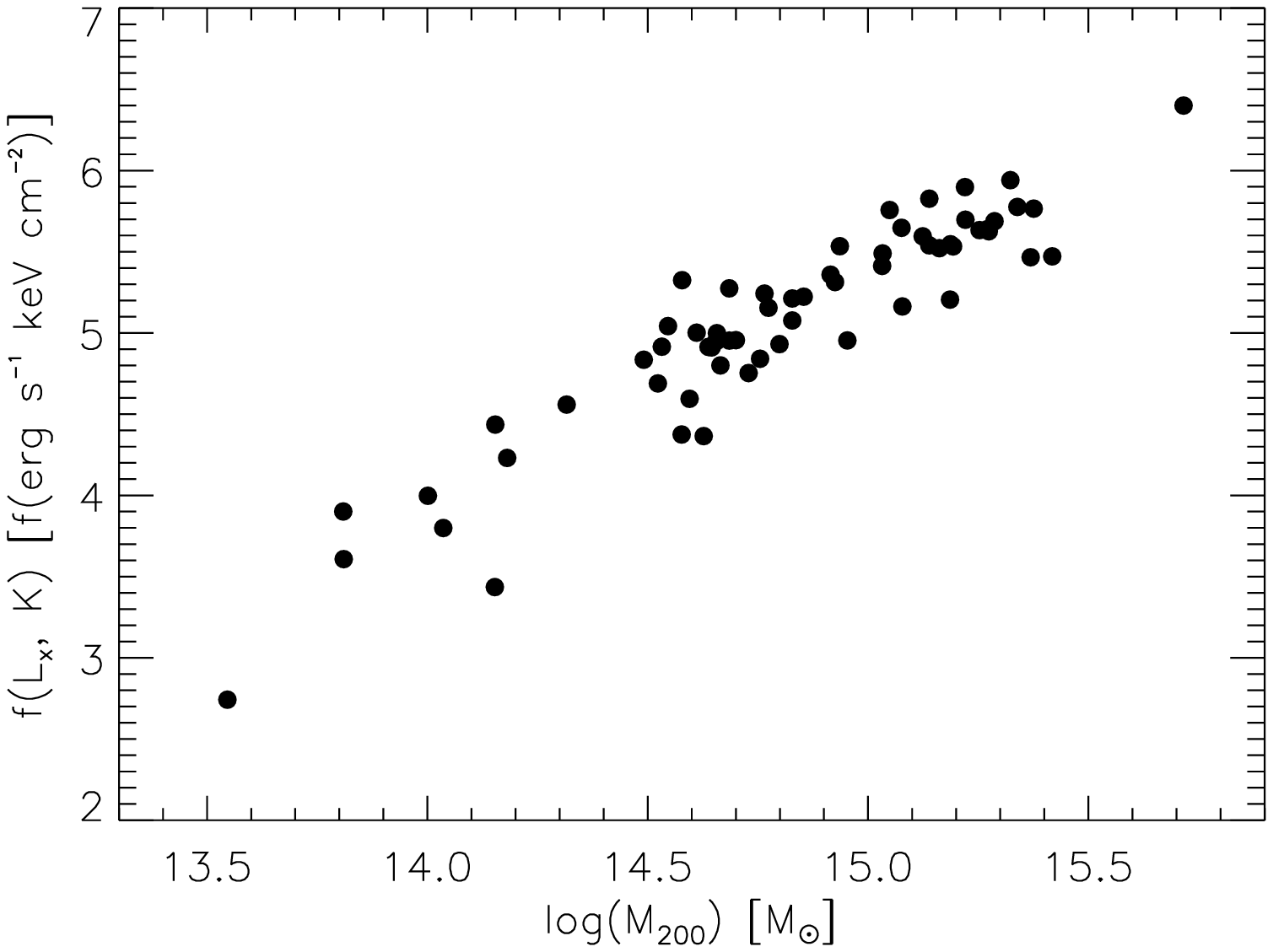}
%
%
\caption{\emph{Left:} $L_{\rm X}$ vs.\ $M_{\rm tot}$ for the \reiprichgcs\ 
clusters (RB02). \emph{Right:} Same as left but all luminosities multiplied by
the central entropy to the power of 0.361, resulting in a reduction of scatter.}
\label{reiprich:EL-M}       
\end{figure}
Having a continuous measure for the ``strength'' of a cool core one can try to
include it as a scaling parameter in the $L_{\rm X}$--$M_{\rm tot}$ relation;
e.g., O'Hara et al.\ \cite{omb06} used the central surface brightness for this
purpose. Here we
play with the central entropy. Figure~\ref{reiprich:EL-M} shows the $L_{\rm
X}$--$M_{\rm tot}$ relation again (left) and then, on the right, the same
relation but all $L_{\rm X}$ values multiplied with the central entropy,
$K^\alpha$, and $\alpha=0.361$ chosen such that scatter is minimized. And,
indeed, such a scaling does reduce the scatter. Again, we will work this out in
more detail once we are completely done with the Chandra analysis. The specific
choice of using central entropy to reduce scatter will possibly only be of
limited practical value because if the data are good enough to determine the
central entropy then $M_{\rm tot}$ is probably better determined directly from
the density and temperature profile than from the $L_{\rm X}$--$M_{\rm tot}$ 
relation.

\section{Summary}
\label{reiprich:summary}

Soon we should be able to quantify robustly the intrinsic scatter in
scaling relations directly from cluster data, eliminating the need to estimate
it indirectly by comparison to other cosmological probes. Even if it turns out
that cool cores cause a relatively large intrinsic scatter, it is
straightforward to correct for the resulting effects in cosmological tests. So,
cool cores do not appear to be a showstopper for using clusters for precision
cosmology.

Something else that will be required in the near future from the X-ray cluster
community is a coordinated effort to perform detailed consistency checks,
similarly to what the weak lensing and simulation communities have already done
\cite{fwb99,hvb06}. We are trying to do a first simple step in this direction by
analyzing the \reiprichgcs\ sample independently with Chandra and XMM-Newton but
a larger scale effort involving several more groups and also simulations is
necessary to convince the general cosmology community that cluster systematics
are sufficiently under control.

%
%


\printindex
\end{document}